\documentclass[reprint,amsmath,amssymb,aps,plr]{revtex4-1}
\usepackage[english] {babel}
\usepackage{hyperref}
\usepackage{amsfonts,graphicx}
\usepackage{hyperref}
\usepackage{subcaption}
\usepackage{epstopdf}

\begin{document}

\title{Current Saturation in Nonmetallic Field Emitters}

\author{Stanislav S. Baturin}%
\affiliation{PSD Enrico Fermi Institute, The University of Chicago, 5640 S. Ellis Ave., Chicago, IL 60637, USA}%
\email{s.s.baturin@gmail.com}%
\author{Alexander V. Zinovev}%
\affiliation{Materials Science Division, Argonne National Laboratory, 9700 S. Cass Ave., Argonne, IL 60439, USA}%
\author{Sergey V. Baryshev}%
\affiliation{Department of Electrical and Computer Engineering, Michigan State University, 428 S. Shaw Ln., East Lansing, MI 48824, USA}%
\email{serbar@msu.edu}%

\begin{abstract}
\noindent 
Field emission from metals was well understood nearly a century ago. Description of its core process given by Fowler and Nordheim was an early success of emerging quantum mechanics. It paved the way to field emission and ion microscopy that were the first methods that imaged metal surfaces at nanometer and atomic resolution. Contrastively, in 1960's and later on it was discovered that nonmetallic (III-V and II-VI semiconductors, diamond, carbon nanotubes, amorphous carbon) field emitters do not obey the very basic Fowler-Nordheim law. In experiments, the output current stops growing with the electric field, and current-voltage characteristic switches from diode-like to resistor-like behavior. This general phenomenon, present in such a broad spectrum of materials, is referred to as current saturation effect and has remained unexplained for more than five decades. Here, we propose a unified and transparent concept that explains the current saturation effect in any nonmetallic field emitter.
\end{abstract}

\maketitle 
\textbf{\textit{Introduction}}
Field emission electron sources are the basis for cutting-edge microscopy \cite{jeol,micros}, X-ray medical devices \cite{10.1109/JMEMS.2014.2332176}, mass spectrometers \cite{10.1007/s13361-015-1212-0}, high power systems \cite{10.1016/S0927-796X9800014-X,10.1109/TED.2009.2015614,10.1103/PhysRevLett.77.2320,BRA,10.1063/1.4901723}, satellite  thrusters (see \cite{Trst} and references therein) and many other applications. In the realm of finding inexpensive and reliable field emission devices, beyond Spindt cathodes \cite{Spt1,Spt2}, that can simplify and scale complex systems packaging via thin film and micro- and nano-fabrication technologies, many novel advanced materials are studied. Main emphasis is being placed on exploring new field emission materials \cite{WChoi,NNS,10.1021/nn201043a} or engineering traditional semiconductor materials \cite{Nat,10.1109/JMEMS.2014.2332176,10.1109/TED.2016.2593905}, or both \cite{10.1002/adfm.201300322}. In other words, advanced device fabrication technologies are focused on materials that are not traditional metals. These nonmetallic materials often have very low turn-on fields that lead to reduced voltage requirements in a system and feature stronger dependence of the output current on the electric field, as compared to traditional metals. At the same time, the early ignition and initially stronger current-electric field dependence of nonmetallic field emitters in low current regime (be it conventional silicon or carbon nanotube, CNT, fibers) gives way to current saturation when the emitter strongly departs from the Fowler-Nordheim (FN) law \cite{1,10.1063/1.117638,18,10,11,12,13,14,15,16}. The existence of current saturation effect was known for many decades and remains an outstanding, unsolved problem. The effect impacts the total output current and therefore remains a serious challenge and obstacle by impeding further improvement of current field emission device technologies and developments toward novel applications.

The current saturation cannot be explained by common space charge effect \cite{17}, because if experimentally determined onset saturation current is normalized by the formal emission area derived from the tip radius of a post/wire/needle like emitter, the current density remains orders of magnitude lower compared to the current density of $\gtrsim10^6$ A/cm$^2$. The current density of $\gtrsim10^6$ A/cm$^2$ has been proved to promote the space charge effect that forces metal field emitters to deviate from the FN law \cite{FNp}. Therefore, the saturation behavior is a general phenomenon specific to nonmetallic field emitters that have an explanation apart from the space charge effect. Other main mechanisms that are likely behind the current saturation effect are (1) surface termination of the emitting tip by foreign molecules that change the structure of the potential barrier for electrons to tunnel through \cite{18}; applied voltage loss (2) along the field emitter or (3) at the interface between the emitter and its supporting bulk substrate/base that can be effectively described by the serial resistor model \cite{9}. One of the main problems with the surface termination mechanism is that, on one hand, it lowers the turn-on field while, contradictory, it limits the output current. Also, molecular adsorbates do not seem to be causing current saturation in metallic emitters. Minoux et al., using simulations and experimental evidence for single CNT \cite{12}, and most recently Forbes, theoretically \cite{19} demonstrated in elegant ways that the voltage loss, in other words significant resistance, along the field emitter post/wire should play a most significant role in the onset of the field emission current saturation. In Ref.\cite{12}, high temperature annealing experiments demonstrated that the improved emitting tip and the entire CNT crystallinity has enhanced the output current such that the saturation onset increased by three orders of magnitude. The presented facts suggest the importance of intrinsic factors (bulk material properties) over extrinsic factors (surface termination and space charge) behind the current saturation and FN law breakdown.

Here, taking the course of considering the mechanism of applied voltage loss along the field emitter due to bulk material properties, we propose a unified concept that explains the basic saturation mechanism, i.e. the nature of the hypothetical serial resistor, and the fundamental difference in the saturation behavior of nonmetallic materials. First, we derive the basic formula. Then, using the formula, we calculate all the current-voltage characteristics and compare against experimental curves for single-tip and arrayed \textit{p}- and \textit{n}-type Si emitters, and for single CNT and CNT fiber emitters. The proposed formalism could become a predictive tool to search for new prospective materials or to optimize existing ones, and could therefore solve the technological aspect of nonmetallic field emitter devices by helping boost their performance.


\textbf{\textit{Theoretical model}}

\begin{figure}[t]
\includegraphics[scale=0.18]{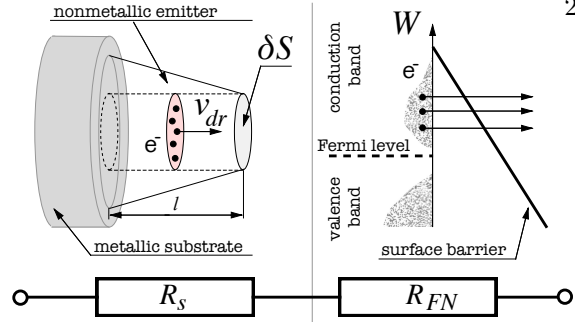}
\caption{A schematic diagram of the proposed cylindrical channel
electron transport and electron field emission.} \label{Fig:1}
\end{figure}

First we consider a nonmetallic layer and consider electron transport from the
substrate to the nonmetal-vacuum boundary. By resolving and
calculating the emission area, we know that emission is limited to
a number of discrete emitting areas across the surface \cite{16}.
So we propose the conductive cylindrical channel concept as
illustrated in Fig.\ref{Fig:1}. Current that is flowing through
the channel we express using well known formula
\begin{align}
I_s=\frac{|e|N}{\Delta t}.
\end{align}
Here $N$ is the number of electrons in the infinitely thin disk of
an area $\delta S$, $e$ is the electron charge and $\Delta t$ is
the time of flight, i.e. time it takes an electron to travel from
the substrate to the surface of the nonmetallic layer. If we
assume that material is isotropic then the number of electrons $N$
can be expressed through the electron volume density $n$ as
\begin{align}
N=n^{2/3} \delta S.
\end{align}
Consequently, the current density can be written in the form
\begin{align}
\label{eq:js}
j_s=\frac{|e|n^{2/3}}{\Delta t}.
\end{align}
Time of flight
$\Delta t$ can be expressed through the drift velocity of
electrons as
\begin{align}
\label{eq:tof}
\Delta t=\frac{l}{v_{dr}(E)}.
\end{align}
Here $l$ is the length of the conducting channel.

It is known \cite{Sze} that drift velocity $v_{dr}$ in a
semiconductor and semiconductor devices \cite{10.1109/16.297726}
depends on the electric field $E$ inside the bulk as
\begin{align}
\label{eq:drsp}
v_{dr}(E_b)=\frac{v_\infty \mu E_b}{(v_\infty^\gamma+\mu^\gamma E_b^\gamma)^{1/\gamma}},
\end{align}
here $v_\infty$ is the saturation drift velocity at high internal
electric field, $\mu$ is the charge carrier mobility and $\gamma$
determines how sharply the drift velocity approaches the
saturation velocity. It was found that $\gamma$ is equal to $2$
and $1$ for electrons and holes, respectively. The saturation
velocity $v_\infty$ can be calculated through the free electron
mass $m_e$ and optical phonon energy $W_{op}$ in traditional
semiconductors as $v_\infty=\sqrt{\frac{8W_{op}}{3\pi m_e}}$  and
it is always very close to $10^7$ cm/s. Typically, the saturation
velocity is an experimentally determined quantity to be used
together with formula \eqref{eq:drsp}. Moreover, regardless what
scattering mechanisms are, even exceptionally high mobility
($\sim$$10^4$ cm$^2$/(V$\cdot$s) materials such as CNT
\cite{24} and graphene
\cite{10.1038/nnano.2008.268} and devices such as 2D electron gas
HEMTs \cite{10.1038/nature10677} also have saturation velocities
close to $10^7$ cm/s.
\begin{figure*}
\includegraphics[width=\textwidth,height=3.5cm]{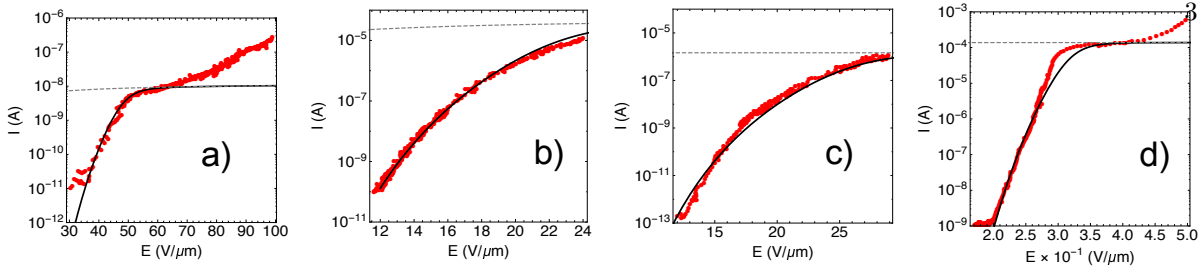}
\caption{Comparison with the experimental results. Red dots are:
a) Ref.\cite{8} single $p$-type Si nano-tip, b) Ref.\cite{22}
array of $n$-type Si nano-tips, c) Ref.\cite{12} single CNT d)
Ref. \cite{10} CNT fiber. Solid lines are formula \eqref{eq:tot}
with corresponding parameters from Table \ref{tb:1}. Dashed lines
are formula \eqref{eq:jsf} with parameters from Table \ref{tb:1}
with $j_s(E)$ multiplied by $S_{av}$.} \label{Fig:2}
\end{figure*}
We assume that if we apply external field $E$ to the surface then
internal field inside the bulk is simply $E_b=E/\varepsilon$,
where $\varepsilon$ is the relative dielectric permittivity. With
this approximation we rewrite \eqref{eq:drsp} as
\begin{align}
\label{eq:drspf}
v_{dr}(E)=\frac{v_\infty \mu E}{(\varepsilon^\gamma v_\infty^\gamma+\mu^\gamma E^\gamma)^{1/\gamma}}.
\end{align}
By combining \eqref{eq:js}, \eqref{eq:tof} and \eqref{eq:drspf} we arrive at
\begin{align}
\label{eq:jsf}
j_s(E)=\frac{|e|n^{2/3}}{l} \frac{v_\infty \mu E}{(\varepsilon^\gamma v_\infty^\gamma+\mu^\gamma E^\gamma)^{1/\gamma}}.
\end{align}
Formula \eqref{eq:jsf} gives estimate of the maximum current
density one can drain from a nonmetallic emitter under external
field. Ultimately in the limit of a very high external field $E
>>\varepsilon v_\infty/\mu$  we have
\begin{align}
\label{eq:jslim}
j_s^{\mathrm{max}}=\frac{|e|n^{2/3}v_\infty}{l}.
\end{align}

Next we consider electron field emission from the surface of the
nonmetallic emitter to the vacuum. Filed emission current density is
usually approximated by FN law \cite{FNp} as
\begin{align}
\label{eq:jFN}
j_{\mathrm{FN}}(E)=a \frac{\beta^2E^2}{\phi} \exp\left(- \frac{b \phi^{3/2}}{\beta E} \right).
\end{align}
Here $E$ is the electric field on the surface, $\beta$ is the
field enhancement factor, $\phi$ is the surface potential barrier
height, and $a=1.54\times10^{-6}$ (A eV V$^{-2}$) and $b=6.83$
(eV$^{-3/2}$ V nm$^{-1}$) are the FN constants.

\begin{table*}[!htb]
\caption{Parameters for the crosscheck with the experimental data} \label{tb:1}
\begin{tabular*}{\textwidth}{@{\extracolsep{\fill}}lccccccccc}
\hline
\hline
  type  &number of&$\phi$ (eV) &$\mu$ (cm$^2$/V$\cdot$s)& $n$ (cm$^{-3}$) & $\varepsilon$ & $v_{\infty}$(cm/s) &$S_{av}$ (cm$^2$) & $l$ (cm)& $\beta$ \\
  &emitters&\\
\hline
 \textit{p}-type Si & 1 \cite{8}                 & 4.7  & 450 \cite{21}       & 3$\times$10$^{15}$    &   12 &  8$\times$10$^{6}$ \cite{21}   &  $\sim$2.82$\times$10$^{-11}$\cite{8}      & 6$\times$10$^{-5}$   \cite{8}         &  95  \\
  \textit{n}-type Si & $\sim$100\cite{22}  &4.7   & 100 \cite{21}     & 5$\times$10$^{19}$     &  12 &  10$^{7}$ \cite{21} &  $\sim$1.26$\times$10$^{-9}$\cite{22}      & 5.7$\times$10$^{-4}$ \cite{22}      & 265 \\
  CNT        & 1\cite{12}                 & 5.0   & 2,000                 & $\sim$3$\times$10$^{19}$                        & 1    &10$^{7}$\cite{24}               &$\sim$2.82$\times$10$^{-11}$\cite{12} & 3$\times$10$^{-4}$   \cite{12} &  260 \\
  CNT fiber& 1 \cite{10}                  & 5.0   & 10,000\cite{26,KKK}  & $\sim$1.5$\times$10$^{19}$      & 10  &10$^{7}$\cite{24}               &$\sim$7.07$\times$10$^{-6}$\cite{10}        & 0.5\cite{10} & 13,500\\
\hline
\hline
\end{tabular*}
\end{table*}

We now consider an equivalent serial resistor model of the
cylindrical channel as it is depicted in Fig.\ref{Fig:1}. Total
current that goes through nonmetal bulk and the potential barrier
on the surface connected in series can be expressed in the terms
of Ohm's law as
\begin{align}
\label{eq:ohml}
\delta I=\frac{U}{R_s+R_{\mathrm{FN}}}.
\end{align}
Here $U$ is an external voltage and $R_s$ is the equivalent
resistance of the bulk and $R_{\mathrm{FN}}$ is the equivalent
resistance of the FN process. Now, if we assume that electric
field screening by the field emission electron current is low, the
field on the nonmetal surface will be approximately independent of
the emission (FN) current. On the other hand, that means that
voltage across the emitter will be approximately $U\approx El$.
The resistance of the emitter and the equivalent FN resistance can
then be expressed through the corresponding current densities as

\begin{align}
\label{eq:res}
R_s(E)&=\frac{E l}{j_s(E) \delta S}, \\
R_{\mathrm{FN}}(E)&=\frac{E l}{j_{\mathrm{FN}}(E) \delta S}.
\end{align}
We note, the formulas \eqref{eq:ohml} and \eqref{eq:res} are
constructed such that they guarantee exact currents $I_s=j_s
\delta S$ and $I_{\mathrm{FN}}=j_\mathrm{FN}\delta S$ in two
limiting cases when $R_s>>R_{\mathrm{FN}}$ and
$R_s<<R_{\mathrm{FN}}$, respectively. With \eqref{eq:ohml} we have
for the total current
\begin{align}
\label{eq:curl} \delta
I(E)=\frac{j_s(E)j_{\mathrm{FN}}(E)}{j_s(E)+j_{\mathrm{FN}}(E)}\delta
S.
\end{align}
Taking into account that the emission area is an electric field
dependent property \cite{16}, we can write the total emission
current measured in experiment as
\begin{align}
\label{eq:tot}
I(E)=\frac{j_s(E)j_{\mathrm{FN}}(E)}{j_s(E)+j_{\mathrm{FN}}(E)}S(E),
\end{align}
here $S(E)$ is the emission area, $j_s(E)$ is given by \eqref{eq:jsf} and $j_\mathrm{FN}(E)$ is given by \eqref{eq:jFN}.

We test the proposed model against four representative experimental result
sets \cite{8,22,12,10}. 
It is clear from the comparison on Fig.\ref{Fig:2}, our model remarkably predicts the
onset kink point when experimental data start deviating from the
FN law, as well as it quantitatively predicts the saturation
current plateau
$I_s^{\mathrm{max}}=\frac{|e|n^{2/3}v_\infty}{l}\delta S$, i.e.
the total current to be taken away from the nonmetallic field
emitter cannot exceed this value. In addition to that, there are
few more consequences of our analysis:

\noindent (1) During photon-assisted field emission experiments
the output current always goes up because the supply term $N$
increases (increased electron supply generated by light).

\noindent (2) During heat-assisted field emission experiments the
output current always goes up because of the additional thermionic
emission mechanism (current is indeed linear in Richardson
coordinates \cite{7, 10.1063/1.4948328}). As the current
saturation plateau grows higher with temperature increasing, it is
predicted to flatten more. The mobility and the saturation
velocity are responsible for this flattening effect because $\mu$
and $v_\infty$ both steadily diminish at temperatures in excess of
the room temperature.

\noindent (3) With all other parameters fixed, the emission
surface area can be calculated.

\textbf{\textit{Cross check with the experiment}}

As far as no data on $S(E)$ dependence
was available we picked two experiments that were performed on
individual emitters, as we were able to accurately estimate the
emission area by using electron micrograph images of those emitter
tips \cite{8,10,12}. In the experiment with arrayed $n$-type Si
emitters \cite{22}, the total emission area was estimated by
taking the number of emitters into account. In the experiment with
a macroscopic CNT fiber emitter \cite{10}, the total emission area
was estimated by using the radius of the fiber reported by the
authors. Estimated emission area values $S_{av}$ along with all
other parameters used in calculations are listed in Table
\ref{tb:1}. Other parameters like drift velocity and carrier mobility were taken from the online data base of the Ioffe institute \cite{21}.  

All experimental data was extracted by digitizing figures of current-voltage curves in the corresponding reference  \cite{8,22,12,10}.   

When plotting theoretical curve all parameters in the final equation were fixed except $\beta$ and $S_{av}$ that were varied slightly to 
achieve the best fit. We note that  
obtained fitting $\beta$-factor values were in close match with
$\beta$-factors or aspect ratio values reported by the authors of
the experiments used for model cross check comparison. 

\textbf{\textit{Conclusion}}

It was shown that the phenomenon of
current saturation in nonmetallic field emitters, in a way that
they stop obeying the Fowler-Nordheim law, has a clear physical
reason. Namely, the output current is saturated/limited by the
maximal number of electrons could be delivered to the emission
point on the surface through the emitter bulk in the direction
perpendicular to the surface, i.e. it is a combination of how many
electrons are available, how fast and how far they have to travel,
how many exit channels on the surface are available at a given
external electric field. Using the simplified and commonly
considered serial resistor model and the fundamental regularities
of charge carrier transport in semiconductors, a unified concept
and mathematical formalism that quantitatively describes the
current saturation phenomenon was proposed. The model demonstrates
excellent agreement with available experimental data and could be
used as a predictive tool to search of new prospective field
emitter materials.

\begin{acknowledgements}

SSB was supported by the U.S. National Science Foundation under
Award No. PHY-1549132, the Center for Bright Beams, and under
Award No. PHY-1535639. AVZ was was supported by the U.S.
Department of Energy, Office of Science, Materials Sciences and
Engineering Division. S.V.B. was supported by funding from the
College of Engineering, Michigan State University, under Global
Impact Initiative.
\end{acknowledgements}

%

\end{document}